\documentclass[twocolumn,prd,superscriptaddress,aps,floats,floatfix,amsmath,amssymb,secnumarabic,nofootinbib]{revtex4} %
\usepackage[dvips,final]{graphicx}
\usepackage{amsmath}
\usepackage{amsfonts}
\usepackage{amssymb}
\usepackage{latexsym}
\usepackage{graphicx}
\usepackage[english]{babel}
\usepackage{multirow}
\usepackage{float}
\usepackage{url}
\usepackage{hyperref}
\usepackage{xspace}
\usepackage{verbatim}

\usepackage{color}
\usepackage{soul}

\newcommand{\be}{\begin{equation}}
\newcommand{\ee}{\end{equation}}
\newcommand{\ba}{\begin{array}}
\newcommand{\ea}{\end{array}}
\newcommand{\bqa}{\begin{eqnarray}}
\newcommand{\eqa}{\end{eqnarray}}

\newcommand{\vl}{\emph{Via Lactea II}\xspace}
\newcommand{\sv}{\langle \sigma v \rangle}

\newcommand{\Al}{$^{26}\!$Al\xspace}
\newcommand{\Ti}{$^{44}$Ti\xspace}
\newcommand{\sss}{\scriptscriptstyle}

\begin{document}

\title{ Interacting dark matter contribution to the Galactic 511 keV gamma ray emission: constraining the morphology with INTEGRAL/SPI observations}
\author{Aaron C. Vincent\footnote{vincenta@hep.physics.mcgill.ca}}
\affiliation{Department of Physics, McGill University,
3600 Rue University, Montr\'eal, Qu\'ebec, Canada H3A 2T8}
\author{Pierrick Martin\footnote{pierrick.martin@obs.ujf-grenoble.fr}}
\affiliation{Institut de Plan\'etologie et d'Astrophysique de Grenoble, BP 53, 38041 Grenoble cedex 9, France}
\author{James M. Cline\footnote{jcline@hep.physics.mcgill.ca}}
\affiliation{Department of Physics, McGill University,
3600 Rue University, Montr\'eal, Qu\'ebec, Canada H3A 2T8}

\begin{abstract}

We compare the full-sky morphology of the 511 keV gamma ray excess
measured by the INTEGRAL/SPI experiment to predictions of models based
on dark matter (DM) scatterings that produce low-energy positrons:
either MeV-scale DM that annihilates directly into $e^+e^-$ pairs, or 
heavy DM that inelastically scatters into an excited state (XDM)
followed by decay into $e^+e^-$ and the ground state.    
By direct comparison to the data, we find that such explanations are consistent
with dark matter halo profiles predicted by numerical many-body
simulations for a Milky Way-like galaxy.  Our results favor an Einasto profile over the cuspier
NFW distribution and exclude decaying dark matter scenarios whose
predicted spatial distribution is too broad. We obtain a good fit
to the shape of the signal using  six fewer degrees of freedom than
previous empirical fits to the 511 keV data.  We find that the
ratio of flux at Earth from the galactic bulge to that of the disk is between
1.9 and 2.4, taking into account that 73\% of the disk contribution
may be attributed to the beta decay of radioactive $^{26}\!$Al.

\end{abstract}
\maketitle
\newpage
% $\sim 1.8$ ph cm$^{-2}$s$^{-1}$
\section{Introduction}

The 511 keV gamma ray line observed by the INTEGRAL/SPI experiment is
consistent with the annihilation of $\sim (1.5 \pm 0.1) \times 10^{43}$
low-energy positrons per second in a region within $\sim 1$ kpc of the
galactic center (GC), in addition to a fainter ($(0.3 \pm 0.2) \times 10^{43}$
$e^+$ s$^{-1}$) disk-like component that extends along the
galactic plane \cite{Knodlseder:2005yq}. The line is mostly due to parapositronium annihilation of thermal or near-thermal positrons \cite{Churazov:2004as,Jean:2005af}.
The absence of $\gamma$ rays from $e^+$ annihilations in flight implies
that the positrons are injected with energies less than $\sim 3$
MeV \cite{Beacom:2005qv}.
No astrophysical source has
been proven to yield such positrons with the required concentrated 
and approximately axially symmetric spatial
distribution.  

Among conventional sources, radioactive ejecta from
stars, supernovae and gamma-ray bursts can produce a large enough rate
of positrons through $\beta^+$ decay, but their spatial distribution
is not sufficiently confined toward the bulge: they predict a
ratio of bulge to disk luminosities $B/D < 0.5$, whereas observations demand
$B/D > 1.4$.  Other proposed mechanisms also suffer from this problem.
In addition, positrons from pair creation near pulsars or from 
$p$-$p$ collisions associated with cosmic rays or the supermassive black hole tend to be too energetic.
Low-mass X-ray binaries have received attention as a possible source,
but these also do not give rise to large enough $B/D$ 
\cite{Bandyopadhyay:2008ts}.  A comprehensive review of these sources and the challenges they face is
given in \cite{Prantzos:2010wi}.

Dark matter (DM) interactions have the potential to explain the
observed excess, either through direct annihilations of light ($\sim$
few MeV) DM particles into $e^+e^-$ pairs \cite{Boehm:2003bt}, or by
the excited dark matter (XDM) mechanism, in which excited states of
heavy DM ($\chi$) are produced  in $\chi$-$\chi$ collisions, with
subsequent decay of the excited state into the ground state and an
$e^+ e^-$ pair  \cite{Finkbeiner:2007kk,Pospelov:2007xh}.  The latter
scenario has the theoretical advantage that the DM mass is relatively
unconstrained, requiring only that the splitting between the ground
and excited states be less than a few MeV.  

XDM as an explanation for the INTEGRAL/SPI 511 keV excess came under greater
scrutiny in recent years after it was proposed
\cite{ArkaniHamed:2008qn} that nonabelian DM models could naturally
have small $\sim$ MeV mass splittings and simultaneously explain additional
recent cosmic ray anomalies  \cite{Adriani:2008zr,Abdo:2009zk} as well
as hints of direct DM detection \cite{Bernabei:2010mq}.  Ref.\
\cite{Chen:2009av} found that it is not possible to get a large enough
rate of positrons for 511 keV emission in the nonabelian models that
require production of {\it two} $e^+e^-$ pairs (one at each
interaction vertex).  However, the original model of 
\cite{Finkbeiner:2007kk} can give a large enough rate 
\cite{Morris:2011dj} since only one such pair need be produced, which
is energetically easier.  Moreover, variant models involving
metastable DM that scatters through a smaller mass gap 
\cite{Chen:2009dm,Cline:2010kv} also give
a large enough rate, and are largely free of threshold velocity issues.

The aforementioned studies focused primarily on matching the overall
rate of positron production, either ignoring morphological 
constraints or estimating them in a rough way.  
Ref.\ \cite{Ascasibar:2005rw} is the only rigorous analysis  with
respect to dark matter models, done at a time when relatively little
data had yet been accumulated.  More recently, ref.\
\cite{Abidin:2010ea} carried out a study of DM predictions for the 511
keV angular profile, but comparing to a previous fit to the observed
shape \cite{Weidenspointner:2007rs} rather than directly to the
data.  

Our purpose in the present work is to improve upon these earlier
papers by testing the DM model shape predictions directly against the
most recent INTEGRAL data. 
We will then examine how these DM models compare to the phenomenological models obtained in previous studies, such as
\cite{Weidenspointner2008,Bouchet:2010dj}, where the 511\,keV celestial signal is represented by analytical shape functions with several free parameters. 
As we will see, an interesting feature of the DM models is that
their predictions depend on far fewer parameters   and they can thus be a more attractive
candidate if they are shown to provide as good a fit as the phenomenological
parametrizations.

In the remainder of the paper, we first present the known sources of positrons in the galaxy, before discussing our procedure for modeling the 511 keV sky in Sections \ref{sec:DM} and \ref{sec:integrals}. We give our main results, along with the details of our fitting procedure, in Section \ref{sec:results} and briefly discuss the implications of this study in Section \ref{sec:discussion}.

\section{Known backgrounds}
\label{sec:positrons}

In order to correctly model the possible contribution to the 511 keV
signal from DM scattering, it is necessary to subtract from the data
the contributions from known sources of low-energy positrons. They can
be produced from $\beta^+$ decay of \Al expelled from massive
stars, as well as  from $^{44}$Ti and $^{56}$Ni produced in supernovae. These contributions should be 
correlated with the stars in the galaxy, thus contributing dominantly
to the disk component of the observed signal.

The contribution of \Al can be more directly assessed than that of the other radio-isotopes.
During \Al decay, the de-excitation of the resulting
$^{26}$Mg nucleus produces a gamma ray signal at an energy of 1809
keV whose magnitude and morphology has also been mapped by
INTEGRAL/SPI \cite{Diehl:2005py}.
Since each decay produces
a positron and an 1809 keV photon, one can unambiguously
determine the fraction of the 511 keV signal originating from
\Al. Ref.\ \cite{Knodlseder:2005yq} showed
that it accounts for roughly half of the disk
component of the 511 keV signal, and we will confirm this.
The contribution of $^{44}$Ti and $^{56}$Ni positrons cannot be evaluated in that way because of their shorter lifetimes. A corollary is that positron escape from supernova and their remnants can be a serious issue, and prevent the determination of positron injection rate directly from the isotopes yields \cite{Chan:1993,Martin:2010hw}. Estimates of the isotopes production in stars and of positron escape fractions suggest that it should make up most of the
remaining disk emissivity \cite{Prantzos:2010wi,Knodlseder:2005yq}. 

\section{Dark Matter Halo Profile}
\label{sec:DM}

Many-body simulations of the formation of galactic halos by collapsing
dark matter particles predict a triaxial halo (see for example
\cite{VeraCiro:2011nb}), which however becomes more approximately
spherical near the galactic center when the effects of baryons are
taken into account \cite{Tissera:2009cm}.  For simplicity we will
consider the halos to be spherically symmetric in most of the present work, although we will show that adding a realistic degree of oblateness does not significantly alter the fit. 
To further constrain the shape of the halo 
we will refer to results of the
\vl simulation \cite{Diemand:2008in}, which modeled the collapse of a
Milky Way-sized ($2\times10^{12}\, M_\odot$) collection of over
$10^9$ particles. We chose \vl because it was specifically geared towards the study of the dark matter halo of the Milky Way.
Among the many known parametrizations of the radial mass-energy density
distribution, two have been especially successful at parametrizing
results of recent simulations. These are the Einasto profile
\begin{equation}
 \rho(r) = \rho_s 
\exp\left({-\left[\frac{2}{\alpha}\left(\frac{r}{r_s}
\right)^\alpha -1\right]}\right)
\label{einastoProfile}
\end{equation} 
and the generalized Navarro-Frenk-White (NFW) profile,
\begin{equation}
 \rho(r) = \rho_s\frac{2^{3-\gamma}}{(r/r_s)^\gamma(1+r/r_s)^{3-\gamma}}.
\label{NFWProfile}
\end{equation} 
In both cases $r$ is the galactocentric radius, while $r_s$, $\alpha$
and $\gamma$ are parameters fit to N-body simulation results. The main
galactic halo of the \vl simulation may be fit to an Einasto profile
with $r_s = 25.7$ kpc and $\alpha = 0.17$, or to an NFW profile with
$r_s = 26.2$ kpc and a central slope of $\gamma = 1.2$
\cite{Kuhlen:2009kx}. The overall density normalization $\rho_s$ can
be computed from the local dark matter density which we take to be
$\rho_\odot = 0.4$ GeV cm$^{-3}$  \cite{Salucci:2010qr} at the sun's position 
$r_\odot = 8.5$ kpc \cite{Kerr:1986hz}.

\section{DM and the 511 keV sky distribution}
\label{sec:integrals}

Although the decaying DM scenario \cite{Picciotto:2004rp} was already
shown to be highly disfavored in refs.\
\cite{Ascasibar:2005rw,Abidin:2010ea}, for completeness we will retest
it in the present work. The flux of 511 keV photons from an $e^+$
produced in the decay of a metastable DM particle $\chi$ of mass
$m_\chi$ is
\begin{equation}
d\Phi = 2(1-0.75f_p)\frac{d\Omega}{4\pi}\int_{l.o.s.} 
\frac{\rho(\ell)}{m_\chi \tau} d\ell
\label{decayIntegral}
\end{equation} 
The integral is along the observer's line of sight parametrized by
$\ell$, $\tau$ is the lifetime, $\rho(\ell)$ is its
position-dependent density and $f_p = 0.967 \pm 0.022$ is the
positronium fraction \cite{Jean:2005af}. It corresponds to the global probability that a given $e^+e^-$ annihilation take place via positronium formation.
The latter can occur in the triplet state ortho-positronium (o-Ps) or the singlet
state para-positronium (p-Ps). To conserve angular momentum, only p-Ps
may decay into two 511 keV photons.

If the positrons are instead produced in a scattering or annihilation event, the observed flux takes a similar form:
\begin{equation}
 d\Phi = 2(1-0.75f_p)\frac{d\Omega}{4\pi}\int_{l.o.s.}\frac{1}{2} 
\frac{\sv \rho^2(\ell)}{m_\chi^2}d\ell
\label{scatterIntegral}
\end{equation}

where $\sv$ is the thermally averaged cross-section for annihilations
or excitations of the DM particles that produce $e^+e^-$ pairs. Henceforth we will use ``scattering'' as shorthand for either XDM scattering or annihilating light DM, since both processes will look like (\ref{scatterIntegral}) to an observer. The
density-squared dependence of this integral means that the observed
flux is much more concentrated in the galactic center
than in the decay case; this is why scattering gives a much better
fit to the observed shape than do decays.

The forms (\ref{decayIntegral},\ref{scatterIntegral}) are only strictly
correct if positrons annihilate close to where they were formed.
Despite recent studies \cite{Higdon:2007fu,Jean:2009zj} the problem of positron transport in the interstellar medium cannot be considered as fully settled. In the absence of strong theoretical and observational constraints, we will assume that positron transport is a small effect
in the present investigation.  We will briefly return to this issue
in Section \ref{sec:discussion}.

Moreover, we have for simplicity assumed that 
$\langle\sigma v\rangle$ in (\ref{scatterIntegral})
is independent of $r$, but this is not a good approximation for all
models.  In particular, for the standard XDM scenarios with a total energy
gap $\delta E>0$ between the ground state and excited state(s), there
is a threshold value for the relative velocity,
$v_t = 2\sqrt{\delta E/m_\chi}$, which appears in the excitation  
cross section as $\sigma v\sim \sigma_0\sqrt{v^2 - v_t^2}$ 
\cite{Finkbeiner:2007kk}.  Because the DM velocity dispersion 
$v_0(r)$ depends
strongly upon $r$ near the galactic center, this factor can then
introduce significant $r$ dependence into the phase-space average
$\langle\sigma v\rangle$.  There are several situations where this is
not important: MeV DM undergoing pure annihilations \cite{Boehm:2003bt,Huh:2007zw}, metastable
XDM models where $\delta E \ll m_e$ or $\delta E < 0$ 
\cite{Chen:2009av,Cline:2010kv}, and standard XDM models where $m_\chi \gtrsim$ TeV,
in which case $v_t$ is small compare to $v_0(r)$.  For XDM models
with $m_\chi \lesssim$ TeV, a more detailed study should be done.

In addition to the dark matter source of positrons, we included a disk
component that models $\beta^+$ emission  from  radioactive isotopes
including \Al and \Ti, whose flux at earth is analogous to eq.\
(\ref{decayIntegral}); the combination $\rho/(m_\chi\tau)$ becomes a
density per unit time $\dot n$ of positron-producing radioactive decays.
We considered two density models for this component. The first is
a Robin young stellar disk (YD) model \cite{Robin:2004qd, Knodlseder:2005yq},
\begin{equation}
 \dot n_{\sss YD}(x,y,z) = \dot n_0 \left[e^{-\left(\frac{a}{R_0}\right)^2} - e^{-\left(\frac{a}{R_i}\right)^2}\right],
\label{youngDisk}
\end{equation} 
with
\begin{equation}
 a^2 = x^2 + y^2 + z^2/\epsilon^2.
\end{equation} 
The fixed disk scale radius is $R_0 = 5$ kpc and the fixed inner disk trucation
radius is $R_i = 3$ kpc. We varied the vertical height scale $z_0 = \epsilon/R_0$
between 50 pc and 140 pc. (Ref.\ \cite{Diehl:2005py} used the 1809 keV
line to fit the \Al distribution to a YD distribution with $z_0 =$ 125 pc.) For
comparison we also took an old disk (OD) model:

\begin{equation}
  \dot n_{\sss OD}(x,y,z) = \dot n_0 \left[e^{-\left(0.25 +\frac{a^2}{R_0^2}\right)^{1/2}} - e^{-\left(0.25 + \frac{a^2}{R_i^2}\right)^{1/2}}\right],
\label{oldDisk}
\end{equation} 
with $R_0 = 2.53$ kpc, $R_i = 1.32$ kpc and a vertical height scale $z_0$ which was varied from 150 to 250 pc. 

\section{Results}
\label{sec:results}
We tested our DM scenario against the INTEGRAL/SPI data by a model-fitting procedure applied to about 8 years of data collected in an energy bin of 5
keV width centered around 511 keV. For this, a model for the sky emission is convolved by the instrument response function and fitted to the data simultaneously to a model for the instrumental background noise in the Ge detectors.  

Our fitting procedure is the same as the one described in section 4.2.1 of
\cite{Knodlseder:2005yq}. The likelihood $L$ of a model assuming a Poisson
distribution of events in each of the $N$ data bins is
\begin{equation}
 L = \prod_{i = 1}^{N} \frac{\lambda_i^{n_i} e^{-\lambda_i}}{n_i!}.
\end{equation} 
$n_i$ is the number of events recorded in bin $i$ by the SPI experiment, and
$\lambda_i = \sum_k\alpha_ks_i^k + b_i(\beta)$ is the predicted number of counts
per bin, including the background $b_i$ and the source $s_i^k =
\sum_jf_j^kR_i^j$. The factor $R_i^j$ is the instrument response matrix and
$f_j^k$ is the intensity computed with the line-of-sight integrals. In our case,
the sum over $k$ has two terms: the dark matter term and the disk component. The
coefficients $\alpha_k$ and $\beta$ are the scaling factors that are adjusted by
the fit. The result of fixing the normalization $\alpha_{DM}$ is to fix
$\left(m_{\chi}\tau_{\chi}\right)^{-1}$ in the case of decay and
$\sv_{\chi}m_{\chi}^{-2}$ for dark matter scattering. We use the maximum likelihood ratio test to estimate detection significances and errors. We calculate the
log-likelihood ratio 
\begin{equation}
   \hbox{MLR} = -2(\ln L_0 - \ln L_1),
\label{MLReq}
\end{equation} 
where $L_1$ is the maximized likelihood of the model being tested, and $L_0$ is
the maximum likelihood of the background model only, \textit{i.e.,} $\alpha_k =
0$. 

We compare the results of our DM models to the phenomenological description by Weidenspointner \textit{et al.} \cite{Weidenspointner2008}, where the authors fitted two spheroidal Gaussians and a young stellar disk to the then-available four-year data set\footnote{The 8 degrees of freedom in the reference model are: the width and normalization of each Gaussian, the inner and outer disk truncation, the disk scale height and the disk normalization.}.
We have updated their analysis, using the currently available eight-year data set and find an MLR of 2693. Although non-nested models cannot be directly compared through the MLR, this serves as a the figure of merit for a model such as the dark matter ones to match, if
it is to provide a competitive fit relative to the phenomenological
shape models.

\begin{figure}[ht]
\hspace{-0.4cm}
\includegraphics[width=0.5\textwidth]{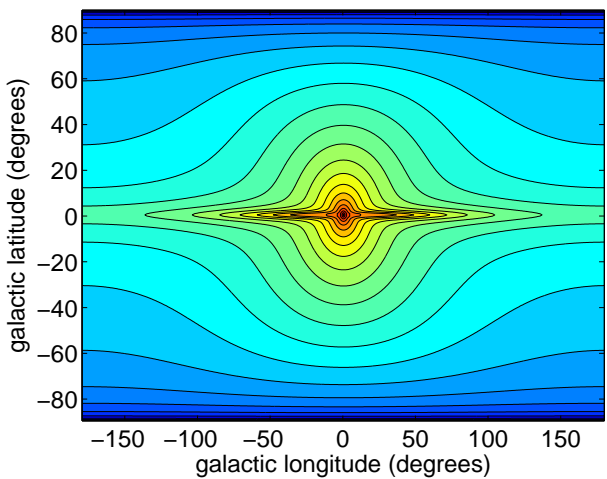}
  \caption{Intensity skymap predicted by \textit{Einasto + disk} model. 
The bulge component is due to emission from scattering or annihilating dark matter in an 
Einasto profile, and the disk component can be attributed to decay of 
radioactive species including mainly \Al.}
\label{skymap}
\end{figure}
\begin{figure}[ht]
\hspace{-0.4cm}
\includegraphics[width=0.5\textwidth]{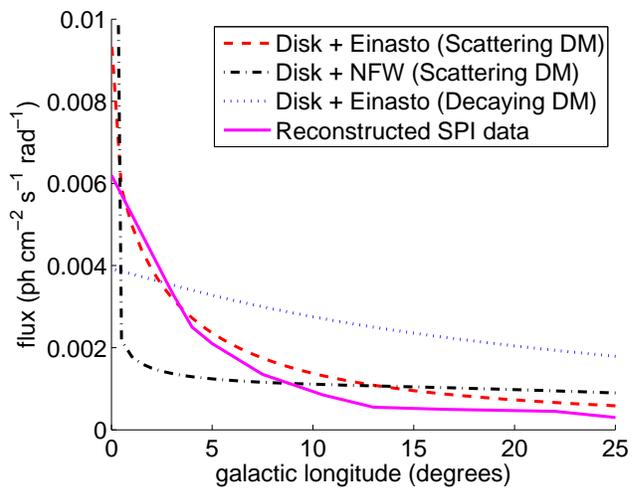}
  \caption{Longitudinal dark matter profiles for the three dark matter models 
considered, including the disk component from radioactive isotopes. Fluxes are 
integrated over galactic latitudes $-15^\circ < b < 15^\circ$. ``Scattering'' refers to either scattering multistate dark matter or annihilating light dark matter. The solid 
magenta line is left-right averaged, reconstructed SPI data from 
\cite{Prantzos:2010wi}, taken from the skymaps of \cite{Weidenspointner:2008zz}.}
\label{profilesfig}
\end{figure}

\begin{table*}[ht]
\vspace{5mm}
% \begin{minipage}[b]{1.8\linewidth}
% \centering
\caption{Summary of best fits to the INTEGRAL/SPI data, with parameters fixed to results of the \vl simulation. This corresponds to $r_s = 26$ kpc and $\alpha = 0.17$ for an Einasto profile (\ref{einastoProfile}) or $\gamma$ = 1.2 for an NFW profile (\ref{NFWProfile}). The disk component is the young disk (\ref{youngDisk}) with $z_0$ = 125 pc. All-sky fluxes are in units of 10$^{-4}$ ph cm$^{-2}$s$^{-1}$, the lifetimes $\tau$ are in seconds, and cross-sections $\sv$ have units of cm$^{3}$ s$^{-1}$. We have highlighted the best fit scenarios in bold. 
}
\begin{ruledtabular}
\begin{tabular}{ l l c c c c }
  \textbf{Channel} & \textbf{Profile} 	& \textbf{MLR} 	 & \textbf{Disk flux}   & \textbf{DM flux} & \textbf{DM lifetime or cross-section} \\ \hline 
  \multirow{2}{*}{decay}	&  Einasto only	&  2139	 & ---      		& 174.5 $\pm$ 3.5  & $\tau_\chi = 1.1 \times 10^{26} $(GeV$/m_\chi)$ \\ 
				&  Einasto + Disk&  2194 & 10.60 $\pm$ 1.42     & 148.6 $\pm$ 5.1  & $\tau_\chi = 1.3 \times 10^{26} $(GeV$/m_\chi)$ \\ \hline
  \multirow{5}{*}{scattering}	&  Einasto only	&  2611  & ---	 		& 24.02 $\pm$ 0.47 & $\sv_\chi = 5.8 \times 10^{-25}(m_\chi/\mathrm{GeV})^2$\\
				&  \textbf{Einasto + Disk}&  \textbf{2668} & \textbf{9.98 $\pm$ 1.32}   & \textbf{21.16 $\pm$ 0.59} & $\boldsymbol{\sv_\chi = 5.1 \times 10^{-25}(m_\chi/\mathrm{GeV})^2}$\\
				& \textbf{Einasto (oblate) + Disk} &   \textbf{2669}	& \textbf{8.74 $\pm$ 1.31} &  \textbf{21.06 $\pm$ 0.61}	& $\boldsymbol{\sv_\chi = 4.9 \times 10^{-25}(m_\chi/\mathrm{GeV})^2}$\\
				&  NFW only    &  1602	 & 	---		& 6.72 $\pm$ 0.17  & $\sv_\chi = 8.2 \times 10^{-26}(m_\chi/\mathrm{GeV})^2$\\
				&  NFW + Disk    &  2155 & 26.45 $\pm$ 1.25     & 4.90 $\pm$ 0.18  & $\sv_\chi = 6.1 \times 10^{-26}(m_\chi/\mathrm{GeV})^2$ \\
\end{tabular}
\label{vlFitTable}
% \end{minipage}
\end{ruledtabular}
\end{table*}

\begin{figure}[ht]
\hspace{-0.4cm}
\includegraphics[width=0.5\textwidth]{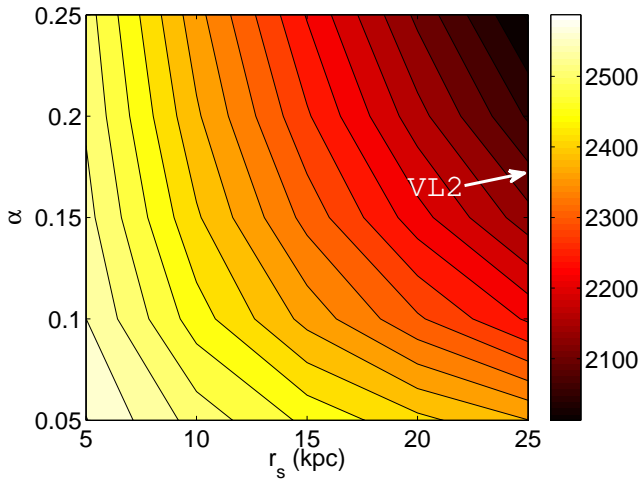}
\caption{Maximum log-likelihood ratio (MLR) obtained in the decaying dark matter + young disk scenario as a function of the Einasto halo parameters. The values favored by the \vl N-body simulation, labeled \textit{VL2}, do not give a good fit to the INTEGRAL/SPI data and are far away from the favored region.}
\label{decayMLR}
\end{figure}

\begin{figure}[!ht]
\hspace{-0.4cm}
\includegraphics[width=0.5\textwidth]{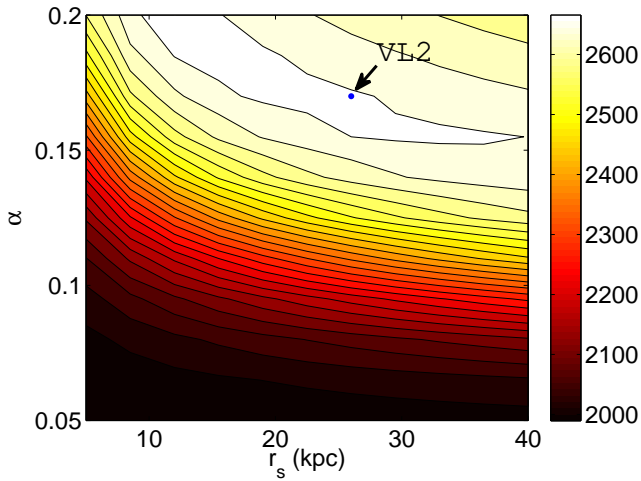}
\caption{Same as Figure \ref{decayMLR}, but with scattering dark matter (\ref{scatterIntegral}). The MLR obtained with the \vl parameters (white dot) is within $\Delta MLR$ = 5 of the best fit ($r_s = 12$ kpc, $\alpha = 0.2$), which means that the VL2 parameters likely correspond to the correct model if the scattering or annihilating dark matter hypothesis is true.}
\label{scatterMLR}
\end{figure}

We performed two analyses, firstly fixing $\alpha$ and $r_s$ to values favored by \vl, using the young disk model
parameters favored by the \Al  analysis of \cite{Diehl:2005py}, and finding the
overall normalizations of the disk and Einasto components that best fit the INTEGRAL/SPI data. As a second analysis, we varied
the parameters $\alpha$ and $r_s$ of the Einasto profile, as well as the height
scales $z_0$ for both young and old disk populations. As we will show, adding
these three extra degrees of freedom does not signifcantly improve the likelihood
of the model, suggesting that the \vl parameters are a good fit for the scattering XDM or annihilating DM
hypothesis. 

Table \ref{vlFitTable} summarizes our main results. The dark matter halo parameters
were set to those favored by \vl, for an Einasto (NFW) profile with  $r_s$ = 26
kpc and $\alpha$ = 0.17 ($\gamma$ = 1.2). We used the young disk model
(\ref{youngDisk}) of \cite{Diehl:2005py}, with the fixed scale height $z_0$ = 125
pc corresponding to the \Al distribution inferred from 1809 keV line data. We
considered both decaying (\ref{decayIntegral}) and scattering
(\ref{scatterIntegral}) dark matter. The scattering scenario provided a
consistently better fit ($\Delta$MLR$> 400$), and the fit to the Einasto profile
was significantly better than to the NFW profile ($\Delta$ MLR$ = 513$). Motivated by the triaxial halo shapes mentioned above \cite{Tissera:2009cm}, we also examined an oblate Einasto profile with a semi-major axis ratio $c/a = 0.8$. This is denoted ``Einasto (oblate) + disk'' in Table \ref{vlFitTable}. While this reduced the required flux from the disk component, it did not produce any significant change in MLR.

The best-fit lifetimes (cross-sections) of the XDM model in the decaying (scattering)
scenario are presented in the final column of Table \ref{vlFitTable}. Figure
\ref{skymap} shows the all-sky map of the {Einasto + disk} best fit to the
INTEGRAL/SPI data, and Figure \ref{profilesfig} shows the longitudinal profile of
the three dark matter models (including disk components) in comparison with a
reconstruction of the SPI data. This clearly illustrates how decaying dark matter
produces a profile that is far too flat, while an NFW distribution results in an
unrealistic sharp central peak. Decaying dark matter in an NFW profile (not
illustrated) displays a combination of these flaws.  On the other hand, the 
scattering model produces MLR $=2668$, which is not far below that of
the best-fit phenomenological
model, the latter having MLR $=2693$ and six additional fitting parameters. The reduced $\chi^2$ of our dark matter model computed on a pointing basis is as good as that of the phenomenological model, with a value of 1.007.

Letting $r_s$, $\alpha$ and $z_0$ vary freely yields some improvement. Figure
\ref{decayMLR} shows a contour plot of the MLR obtained from the decay scenario
(\ref{decayIntegral}). The favored region in the lower-left corner, with an MLR
of 2558, corresponds to an extremely cuspy DM halo that is quite far removed
from realistic DM halo models. 

The equivalent picture for scattering DM is illustrated in Figure
\ref{scatterMLR}. The overall best fit was found to be for a profile with $\alpha
= 0.2$, $r_s = 12$ kpc and $z_0 = 140$ pc, with an MLR of 2673. However, this
difference is only marginally significant. Indeed, by adding three degrees of
freedom, such an improvement should happen by chance 17\% of the time due to
statistical fluctuations in the data.
We found that the young disk (YD) model consistently gave a better fit
than the old disk (OD) model, and that adjusting $z_0$ over a range from 70 to
200 pc did not produce any significant improvement in the MLR. 
Finally, we checked that choosing a closer value for the galactocentric distance of $R_\cdot = 8.2$ kpc, as suggested by recent studes such as \cite{Salucci:2010qr} produced a negligible change in the fit ($\Delta MLR < 1$). 

\section{Discussion and Conclusion}
\label{sec:discussion}

We have made the first direct comparison of dark matter predictions for the
observed 511 keV spatial intensity distribution since the earliest data release
of INTEGRAL/SPI.  
Our favored fit corresponds to a scattering excited DM or annihilating light DM model in an Einasto density
distribtion (\ref{einastoProfile}) with parameters fixed to the \vl results.
We confirm previous analyses showing that decaying dark matter is ruled out
due to its too-broad spatial distribution.
After correct normalization of the intensity, our best-fit model requires a 
cross section for $\chi\chi$ to produce positrons 
 of $\sv_\chi = 5.1 \times 10^{-25}(m_\chi/\mathrm{GeV})^2$
cm$^{3}$s$^{-1}$. If $m_\chi$ is in the 10-1000 GeV range as favored by most WIMP
models, this means $\sv$ is in the interval 
 $\left[ 10^{-23},  10^{-19}\right]$ cm$^{3}$s$^{-1}$.  The fact that this is
far above the annihilation cross section of $3\times 10^{-26}$ cm$^{3}$s$^{-1}$
needed to get the observed relic density is not problematic, because the physical process
required in these models is inelastic scattering to an excited state rather than
annihilation. 
 
Because we neglected $r$-dependence in the averaged cross section
$\langle\sigma v\rangle$, these results apply to upscattering XDM with 
high masses $m_\chi \gtrsim$ a few TeV, metastable XDM models
\cite{Chen:2009dm,Cline:2010kv}, and direct annihilation of MeV DM. 
To cover the case of lighter XDM models, a more detailed analysis
taking account of the radial dependence of the DM velocity dispersion
in the Galaxy would be needed.  We hope to return to this in future
work.

For light $\sim$ MeV DM annihilating directly into $e^+ e^-$, our
required cross section is $\sv\sim 10^{-31}$ cm$^{3}$s$^{-1}$, which
is too small to give the right relic density.  This need not be
a problem; it only requires there to be additional stronger
annihilation  channels into invisible particles, for example 
dark gauge bosons \cite{Huh:2007zw} or dark neutrinos \cite{Cline:2011uu}.

There are two unknowns that could change our analysis in significant ways. One is
the distance by which positrons propagate between creation and annihilation. 
If it is larger than $\sim 100$ pc, it could alter the overall breadth of the spatial extent of the signal, as well as
introduce deviations from axial symmetry,  depending on the conditions of the interstellar medium in the bulge.
Further observational evidence constraining the
structure of magnetic fields (for example synchrotron emission studies
\cite{Bringmann:2011py}) will be needed to reduce
these uncertainties. A second unknown is the degree of departure of the DM halo
from spherical symmetry, which definitely occurs in $N$-body simulations \cite{Tissera:2009cm}. We showed that adding some oblateness had little effect on the fits, though the
nature and extent of triaxiality near the galactic center depends heavily upon
the inclusion of baryons in the simulations, a challenging field which is still
in its early stages.  We look forward to improvements in these studies that will
help to constrain the theoretically expected extent of triaxiality in the DM halo.

We have confirmed the findings of previous studies concerning the disk emission. Given
a young disk model for the distribution of \Al, the observed flux of 1809 keV
gamma rays \cite{Diehl:2005py} translates into an expected 511 keV flux of (7.33
$\pm$ 0.89) $\times 10^{-4}$ ph cm$^{-2}$s$^{-1}$. This alone accounts for 73\%
of the disk component favored by our model. If similar amounts of \Ti are present
in the Galaxy, there is no need for an extra component to explain the disk
component of the 511 keV signal.   On the other hand, simulations show that in addition to
the DM halo, there may also be a DM disk.  This would give an extra DM
contribution to the disk component of the 511 keV emission.  However, there is as
yet no direct evidence for a DM disk in our own galaxy
\cite{Bidin:2010rj,Pestana:2010hb}.

It is worth emphasizing that only two degrees of freedom were required to obtain
the MLR of 2668 in the DM scattering/annihilation scenario. This is in contrast
to the 8 d.o.f. necessary to obtain an MLR of 2693 with one best-fit phenomenological model. A further advantage of the DM model is
that it is motivated by particle physics and cosmology,  and it has a concrete,
calculable production mechansim for the excess electron-positron pairs. Our 
results are independent of the details of the DM model, so long as the scattering
events lead directly to an $e^+e^-$ pair.

We find these results to be encouraging for the dark matter interpretation of the
511 keV excess, an anomaly that was first seen in 1972 by balloon-borne
detectors \cite{balloon}.  We hope that the experimental hard X-ray / soft gamma-ray astronomy community will be
motivated to consider a higher-resolution instrument that would be sensitive to
the 511 keV region of the spectrum in the future.  Such observations would help
to shed more light on this intriguing possibility, which could be the first 
evidence for nongravitational interactions of dark matter.
\\

\section*{Acknowledgments} 

We thank the anonymous referee for insightful comments that helped to improve our presentation. We would like to thank Evan Mcdonough for his contributions to our skymap models. JC is supported by NSERC (Canada). PM acknowledges support from the European Community via contract ERC-StG-200911, and ACV is supported by an NSERC Alexander Graham Bell Canada Graduate Scholarship.

\bibliography{spibliography.bib}
 \bibliographystyle{h-physrev}
\end{document}